\documentclass[aps,prl,twocolumn,groupedaddress]{revtex4}

\usepackage{graphicx}

\begin{document}
 

\title{Route to nonlocality and observation of accessible solitons}


\author{Claudio Conti, Marco Peccianti, Gaetano Assanto}
\affiliation{
NOOEL - Nonlinear Optics and Optoelectronics Lab,
National Institute for the Physics of Matter, INFM - Roma Tre,
Via della Vasca Navale 84, 00146 Rome - Italy}
\homepage{http://optow.ele.uniroma3.it/opto_2002.shtml}
\date{\today}
\begin{abstract}
We develop a general theory of spatial solitons in a liquid crystalline medium exhibiting a nonlinearity with an arbitrary degree of effective nonlocality. 
The model accounts the observability of \textit{accessible solitons} and establishes an important link with parametric solitons.
\end{abstract}
\maketitle
In various areas of applied nonlinear science, nonlocality plays a relevant
role and radically affects the underlying physics.  Some striking evidences are found
in plasma physics,\cite{Litvak78,Pecseli80,Turitsyn85}
  or in Bose-Einstein condensates (BEC),
\cite{Parola98,Perez00}
 where, contrary to the prediction of purely local nonlinear models, nonlocality may give rise to, or prevent, the collapse of a (plasma- or
matter-)wave.
In nonlinear-optics, particularly when dealing with self-localization and solitary waves, nonlocality is often associated
to time-domain phenomena through a retarded response
(see e.g. \cite{Vanin94,Akhmediev98}); 
spatially nonlocal effects have been associated 
to photorefractive \cite{Durin93,Mamaev97,Abe98,DelRe01} 
and thermal or diffusive responses \cite{Dreischuh99,Warenghem02}. 
To assess the role of nonlocality, theoretical studies 
tend to distinguish between 
{\it highly} and {\it weakly} nonlocal behaviors
\cite{Snyder97,Mitchell99,Krolikowski00}, by comparing 
the spatial extent of the material response (the so-called kernel-function)
 and the optical beam waist.
Specific kernel functions, however, strongly depend on the physical system and, as in the case of BEC, \cite{Perez00} they are hard to determine and apply to experimental results.
On the other hand, they are at the basis of the theory of spatial optical solitons (SOS)
in highly nonlocal media.
\\
 SOS have become the subject of intense theoretical and experimental investigations, both on the grounds of their packet nature and in view of applications, particularly in the exploitation of their wave-guiding character. \cite{Boardman01}  Diverse material properties have been studied in conjunction with SOS existence and properties, including various mechanisms able to counteract diffraction in one or both transverse dimensions. \cite{Stegeman00,Trillo01}  
Spatial solitons due to a local nonlinearity have been known since the original work of Chiao, Garmire and Townes with reference to Kerr media.
 \cite{Chiao64}
In 1997 Snyder and Mitchell \cite{Snyder97} investigated SOS in a highly nonlocal system, i. e., a medium exhibiting a power - rather than intensity - dependent 
refractive index. They introduced the term \textit{accessible solitons} for those spatial solitary waves, owing to the simplicity of the theory and transverse profiles obeying the 2-dimensional equation of a quantum harmonic oscillator (which gives a gaussian profile like the so-called ``gaussons'' \cite{gaussons78}). 
Shen pointed out to this connection as an
intriguing one between distinct fields of modern physics and, provided the required large correlation lengths could be made available, to the demonstration of \textit{accessible solitons} as a challenge well worth undertaking. \cite{Shen97}  
\\
In this Letter we introduce a model able to describe optical spatial solitons and the smooth transition from the purely local (in the limit of a Kerr nonlinearity) 
to the entirely nonlocal case. While pursuing a general theory, 
however, we chose to address a specific and available nonlocal system, i. e., nematic liquid crystals (NLC) in a planar cell.  
NLC have been proven to exhibit a substantial nonlocal nonlinearity of molecular origin \cite{Khoo95} and to support (2+1)-dimensional spatial solitons, 
\cite{Peccianti00} even in the case of spatially incoherent excitations. \cite{Peccianti02} 
After deriving the ruling equations and defining a suitable nonlocal parameter able to span the soliton-family 
from pure-Kerr or Townes-like (T-) to higly nonlocal or \textit{accessible} (A-) solitons, 
we will outline the rather remarkable connection between 
our model and the equations describing quadratic two-color solitons (or \textit{simultons}) in parametric media. \cite{Karamzin76}
\\
Let us consider the simple geometry sketched in figure 1: a planar glass cell containing an undoped NLC with a preset orientation of its molecular director. The aligned liquid crystal, anchored at the bounding interfaces, behaves as a positive uniaxial with $n_{//}$=$n_{\hat{z}}$ 
and $n_{\bot}$=$n_{\hat{x}}$. 
In the presence of an external quasi-static (electric or magnetic) field or special anchoring at the interfaces, the refractive index $n(\theta)$ in the (x, z) principal plane can exhibit a distribution along $x$, with $n_\bot < n(\theta(x)) < n_{//}$. \cite{Tabirian86}
For a light beam linearly polarized along $x$ and propagating along z, with transverse 
size well below the cell thickness L, neglecting 
vectorial effects and adopting the paraxial approximation, the evolution of the optical envelope $A$ is described by the Foch-Leontovich equation:
\begin{equation}
  2ik\frac{\partial A}{\partial Z}+\nabla^2_{XY}A+k_0^2 n_a^2[sin(\theta)^2-sin(\theta_0)^2]A=0
\end{equation}
where $k_0$ is the vacuum wavenumber, $n_a^2$=$n^2_{//}$ - $n^2_{\bot}$ the optical anisotropy,
$k^2=k_0^2(n^2_{\bot}+ n_a^2 sin(\theta_0^2)^2)$, $\theta$ the tilt angle of the NLC director, and $\theta_0$ the tilt in the absence of a light beam.
When an external electric field applied along $x$ and an optical excitation as in equation (1) are present, $\theta$ is subject to re-orientation according to \cite{Khoo95}
\begin{equation}
  \label{eq:angle1}
  K\frac{\partial^2\theta}{\partial Z^2}+K\nabla^2_{XY}\theta+
\frac{\Delta\epsilon_{RF}E^2}{2}sin(2\theta)+
\frac{\epsilon_0 n_a^2 |A|^2}{4}sin(2\theta)=0
\end{equation}
with $K$ the relevant elastic constant taken equal for splay, bend and twist, 
$\Delta\epsilon_{RF}$ the low-frequency anisotropy, and $E$ the rms value of the quasi-static field.
\\
In the absence of a light wave, therefore, the orientation angle $\hat{\theta}$ is determined exclusively by $E$ and,
due to simmetry, only depends on X:\cite{Khoo95}
\begin{equation}
  \label{eq:angle2}
  K\frac{d^2\hat{\theta}}{d X^2}+\frac{\Delta \epsilon_{RF}E^2}{2}sin(2\hat{\theta})=0\text{.}
\end{equation}
In our case the boundary conditions correspond to the planar alignment: 
$\theta(X=-L/2)=\theta(X=L/2)=0$.
In the general case, the angle distribution can be written as: 
\begin{equation}
  \label{eq:angle3}
  \theta(X,Y,Z)=\hat{\theta}(X)+\frac{\hat{\theta}(X)}{\theta_0}\Psi(X,Y,Z)\text{.}
\end{equation}
Taking the cell much larger than the beam waist, 
we can use (\ref{eq:angle2}) 
and (\ref{eq:angle3}) in (\ref{eq:angle1}) and neglect the derivative 
of $\hat{\theta}$. For $\hat{\theta}\cong\theta_0$, at the first order in $\Psi$ we obtain: 
\begin{equation}
  \label{eq:both}
  \begin{array}{l}
2ik\frac{\partial A}{\partial Z}+\nabla_{XY}^2 A+k_0^2 n_a^2\Psi A=0\\
  K\nabla^2_{XYZ} \Psi-\frac{2\Delta\epsilon_{RF}E^2}{\pi}\Psi+
\frac{\epsilon_0 n_a^2}{4}|A|^2=0\text{,}
\end{array}
\end{equation}
having chosen $\theta_0=\pi/4$ in order to maximize the nonlinear response. \cite{Tabirian86}
We write (\ref{eq:both}) in a dimensionless form
 by setting
$A=(A_c/\alpha)a(R/R_c\sqrt{\alpha},Z/\alpha Z_c)exp(iZ/\alpha Z_c)$, 
$\Psi=(\Psi_c/\alpha)\psi(R/R_c\sqrt{\alpha},Z/\alpha Z_c)$
with $A_c^2=8 \Delta\epsilon_{RF}^2 E^4/\pi^2  \epsilon_0 k_0^2 n_a^4 K$,
$Z_c=2k R_c^2$,$R_c^2=\pi K/2 \Delta\epsilon_{RF} E^2$,
$\Psi_c=2 \Delta\epsilon_{RF} E^2/\pi  k_0^2 n_a^2 K$,
and $\epsilon=\Delta\epsilon_{RF} E^2/2\pi k^2 K$.
$\alpha$ is a free parameter, to be used hereafter to trace the family of spatial solitary waves, and 
$(x,y,z)=(X/\sqrt{\alpha}R_c,Y/\sqrt{\alpha}R_c,Z/\alpha Z_c)$ are normalized
coordinates.
The resulting system is:
\begin{equation}
\label{eq:bothadimensional}
\begin{array}{l}
i\frac{\partial\,a}{\partial\,z}+\nabla^2a-a+a\psi=0\\
\frac{\epsilon}{\alpha}\frac{\partial^2\psi}{\partial\,z^2}+\nabla^2\psi-\alpha\psi+\frac{1}{2}|a|^2=0\text{.}
\end{array}
\end{equation}
To underline the physical meaning of the parameter $\alpha$, 
let us consider the case $\epsilon=0$
 and formally write the solution of the re-orientation equation as
$\psi=(1-\nabla^2/\alpha)^{-1}|a|^2/(2\alpha)$. 
For large $\alpha$, we obtain for the field envelope:
\begin{equation}
\label{eq:NLSmodified}
i\frac{\partial\,a}{\partial\,z}+\nabla^2a+a(1+\frac{\nabla^2}{\alpha})
\frac{|a|^2}{2\alpha}=0\text{.}
\end{equation}
Equation (\ref{eq:NLSmodified}) rules collapse-free weakly-nonlocal media,\cite{Turitsyn85},
while the neglecting of terms such as $O(1/\alpha^2)$ 
describes light propagation in Kerr media which, on the contrary, are subject to
catastrophic self-focusing.  
Reducing $\alpha$ increases
 the \textit{degree of nonlocality} in the interaction between medium and optical beam. 
 In the following we will employ $\alpha$ as an arbitrary parameter spanning the
whole family of SOS; we expect large $\alpha$ values to be associated to
T-solitons, whereas small $\alpha$ will address A-solitons. 
Note that, according to \cite{Turitsyn85},  when $\epsilon=0$ the fundamental solitary wave solutions of Eqs.(\ref{eq:bothadimensional}) are stable because they realize an absolute minimum of the Hamiltonian.

Solitary solutions of (\ref{eq:bothadimensional}) are defined by $\partial_z=0$.  Without loss of generality, taking $a$ real-valued we have:
\begin{equation}
  \label{eq:chi2solitons}
\begin{array}{l}
\nabla^2a-a+a\psi=0\\
\nabla^2\psi-\alpha\psi+\frac{a^2}{2}=0\text{.}
\end{array}  
\end{equation}
Noteworthy enough, system (\ref{eq:chi2solitons}) is identical to what determines the 
profile of parametric spatial solitary waves in $\chi^{(2)}$-media. \cite{ Trillo01, Karamzin76} 
This makes an unexpected connection between self-trapped beams in 
two distinct physical systems encompassing rather diverse nonlinear optical responses: the ultrafast electronic 
nonlinearity of quadratic crystals and the slow molecular re-orientation of liquid crystals. 
The similarity is only limited to the profiles of the solitary waves, 
while their dynamic properties, such as stability, are quite different.
As shown below in a practical case, to a first approximation $\epsilon$ is negligible, 
thus the angle ``adiabatically-follows'' the light beam; 
in general this is not true for the harmonic-field of a parametric soliton.
It is well known, in fact, that for small $\alpha$ the $\psi$-field 
(the second-harmonic for $\chi^{(2)}$ crystals, and the re-orientation for NLC) 
is much wider than the $a$-field (see Torruellas {\it et al.} 
in \cite{Trillo01}). 
The opposite holds true 
for $\alpha\rightarrow\infty$: in the $\chi^{(2)}$-SOS literature, this 
is the \textit{Kerr-limit}, its dynamics resembling that of $\chi^{(3)}$ materials. 
In NLC, for optical beams much wider than the re-orientation profile,
T-solitons approximate well the solution (in the framework 
of the validity of the large cell approximation). 
The angle-profile perturbation is localized close to
the optical beam axis, as typically pointed out when addressing the Kerr-like response of NLC.
\cite{Tabirian86,Khoo95}
Here we are rather interested in the opposite limit.
\begin{figure}
\includegraphics[height=40mm]{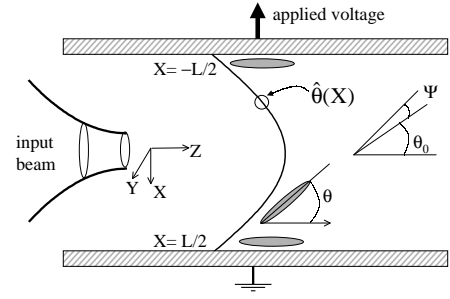}
\caption{Planarly aligned nematic liquid crystal cell for the observation
of spatial solitons. $\hat{\theta}(X)$ is the low-frequency voltage-induced angle displacement. $\theta_0=\hat{\theta}(0)$ and $\Psi$ is the perturbation
due to the propagating optical field.
\label{figure1}}
\end{figure}
\begin{figure}
\includegraphics[width=70mm]{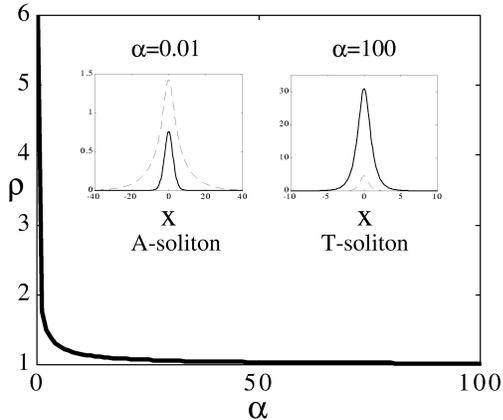}
\caption{
Ratio $\rho$ between the waists of the angle $\psi$ and the field $a$ 
from eq.(\ref{eq:chi2solitons}).
The inserts show two different profiles (dashed line: $\psi$, 
solid line: $a$) for values of $\alpha$ addressing
the two limits in the family of solitary waves.
\label{figure2}}
\end{figure}
\\In figure \ref{figure2} we graph the numerical solutions
 of eqs. (\ref{eq:bothadimensional}) as obtained by a relaxation procedure. 
The ratio $\rho$ between SOS beam- and angle- waists 
(standard deviation) is plotted versus $\alpha$. 
When $\alpha$ approaches zero, i. e., for a response-length extending well beyond the optical waist, the $\psi$-field is much wider than the $a$-field. 
This is the high-nonlocality regime.
In the figure we indicate the two opposite limits, namely A- and T-solitons. 
In the A-limit, however, it is worth proving that eqs. (\ref{eq:bothadimensional}) 
reduce to the model in \cite{Snyder97}. 
The authors in \cite{Snyder97} postulated a highly nonlocal medium in which the index of refraction 
could be expressed as $n^2=n_0^2-\alpha_0^2(\mathcal{P}) R^2$, with $\mathcal{P}$ the optical power. 
 Here we show that the NLC re-orientational nonlinearity does indeed exhibit such feature, as speculated rather skeptically in \cite{Shen97}.

To study the highly non-local regime ($\alpha\rightarrow0$), first we solve eqs. (\ref{eq:chi2solitons}), in the
regions around $r=0$ ($r=\sqrt{x^2+y^2}$) and $r\rightarrow\infty$, and then
match the resulting expressions. 
For $r\cong0$ we introduce the expanded variables \cite{Nayfeh93}
 $\zeta=r/\alpha^{1/2}$ and $\xi=r/\alpha^{1/4}$. Next, we express 
the field as $a=\frac{\hat{a}}{\sqrt{\alpha}}f(\xi)$ with $f(0)=1$,
while $\psi=\psi^{(0)}(\zeta)+(1/\sqrt{\alpha})\psi^{(-1)}(\zeta)$.
At order $O(1/\alpha^{1/2})$ we obtain for $\psi$:
\begin{equation}
  \label{eq:O1}
  \psi_{\zeta\zeta}^{(0)}+(1/\zeta)\psi_{\zeta}^{(0)}+(1/2)\hat{a}^2 f^2(\alpha^{1/4}\zeta)=0
\end{equation}
For $\alpha\rightarrow0$ with $\zeta$ fixed, the solution reads 
$\psi^{(0)}=\psi_0-\hat{a}^2 r^2/8\alpha=\psi_0-\hat{a}^2 \xi^2/8\sqrt{\alpha}$. 
At order $O(1/\alpha^{3/2})$ we have $\psi_{\zeta}^{(-1)}=0$; 
at order $O(1/\alpha^{1/2})$, the field equation reduces to the harmonic
oscillator equation in $f$: 
\begin{equation}
  \label{eq:O2}
  f_{\xi \xi}+(1/\xi)f_{\xi}+\psi^{(-1)}f-(1/8)\hat{a}^2 \xi^2 f=0
\end{equation}
corresponding to the result in \cite{Snyder97}, and $\psi_0=1$ at order $O(1)$.
After some algebra, the perturbative approach provides the SOS profile 
near the origin ($\psi^{(-1)}=2/\xi_0^2$):
\begin{equation}
  \label{eq:O3}
\begin{array}{l}
  a=\frac{2\sqrt{2}}{\sqrt{\alpha} \xi_0^2} e^{-\frac{r^2}{2\sqrt{\alpha} \xi_0^2}}\\
\psi=1+\frac{2}{\sqrt{\alpha}\xi_0^2}-\frac{r^2}{\alpha \xi_0^4}
\end{array}
\end{equation}
with $\xi_0$ an arbitrary parameter. The appearance of another free parameter reflects that, when $\alpha=0$, eqs. (\ref{eq:bothadimensional}) are invariant with respect to 
the transformation $(x,y)\rightarrow(x/\mu,y/\mu)$,$a\rightarrow \mu a$,$\psi\rightarrow 
1-\mu +\mu \psi$, with $\mu$ arbitrary.
The physical meaning of (\ref{eq:O3}) can be elucidated by expressing them in terms of the normalized beam
power $P=\int\int a^2 dx dy$: $a=(\sqrt{2} P/4\pi)exp(-r^2 P/16\pi)$, 
and $\psi=1+(P/4\pi)-(P/8 \pi)r^2$. The latter results can be also 
obtained by using a multiple scales expansion in $\sqrt{P}r$, $Pr$, $...$.
For the field intensity in the original variables($R=\sqrt{X^2+Y^2}$), we have:  
\begin{equation}
  \label{eq:IintermsofP}
  \mathcal{I}=\frac{|A|^2}{2\eta}=
\frac{\mathcal{P}^2}{\pi R_c^2 \mathcal{P}_c}exp(-\frac{R^2}{R_c^2}\frac{\mathcal{P}}{\mathcal{P}_c})
\end{equation}
In (\ref{eq:IintermsofP}) $Z_0$ and $\eta=Z_0/n$ are vacuum and medium impedances, respectively, $c$ the speed of light in vacuum, $n=k/k_0=\sqrt{n_\bot+n_a^2 /2}$ when $\theta_0=\pi/4$, and $\mathcal{P}_c$ 
a reference power which depends on material properties and cell-polarization: 
$\mathcal{P}_c=16 c n \Delta\epsilon_{RF} E^2/k_0^2 n_a^4$.
The soliton profile is gaussian and completely determined by $\mathcal{P}$. Rendering explicit the relation between $\mathcal{P}$ and the (intensity) waist
 $\mathcal{W}$, its existence curve is:
\begin{equation}
  \label{eq:existencecurve}
  \frac{\mathcal{P}}{\mathcal{P}_c}=\frac{ R_c^2}{\mathcal{W}^2}\text{.}
\end{equation}

Let us now investigate the region of large $r$. According to eqs. (\ref{eq:bothadimensional}) the asymptotic behavior of $\psi$ as
 $r \rightarrow \infty$
 is given by the modified Bessel function: $\psi \rightarrow G K_0(\sqrt{\alpha}r)$, governing the angle decay far from the beam axis ($G$ is a constant to be determined).
The angle in (\ref{eq:O3}) and its derivative must match the 
expression at infinity, providing the {\it turning point} $r_T$
 between the two regions: in the presence ($r\cong0$) and in the absence ($r\rightarrow\infty$) of the optical excitation, respectively.
We end up with
\begin{equation}
  \label{eq:turning}
  1+\frac{P}{4\pi}-r_T^2(\frac{P}{8\pi})^2=(\frac{P}{8 \pi})^2
\frac{2r_T K_0(\sqrt{\alpha} r)}{\sqrt{\alpha} K_1(\sqrt{\alpha} r)}\text{,}  
\end{equation}
which can be further simplified by taking into account that, for large arguments, the Bessel functions $K_0$ and
$K_1$ have the same asymptotic behavior. 
The approximated solution is $r_T=\sqrt{\alpha}(8\pi/P)$; being $\mathcal{R}_T=\sqrt{\alpha}R_c r_T$ it reads:
$\mathcal{R}_T /\mathcal{R}_c=\mathcal{P}_c/\mathcal{P}$. 
When $\mathcal{P}>>\mathcal{P}_c$ the profile of the angle is
dominated by the modified Bessel function, and decays with typical length $R_c$.
Therefore, it seems natural to label  ``highly nonlocal'' a regime
when the angle profile is much larger than the beam waist, such that 
  $\mathcal{R}_c>>\mathcal{W}$.
When the power is much greater than $\mathcal{P}_c$ or, 
equivalently, when the soliton waist is much smaller that the extent $\mathcal{R}_c$ of the elastic response, we are in the A-soliton regime.
The power dependent perturbation of the refractive index is:
\begin{equation}
  \label{eq:SMlaw}
  \Delta n(R,\mathcal{P})=
  n_c (\frac{2\mathcal{P}}{\mathcal{P}_c}-
\frac{R^2}{R^2_c}\frac{\mathcal{P}^2}{\mathcal{P}_c^2})\text{,}
\end{equation}
with $n_c=\Delta\epsilon_{RF}E^2/\pi n k_0^2 K$.
Moreover,
 our approach
enables to go beyond the harmonic oscillator.
 By introducing the new transverse scale $\sigma=r/\alpha^{3/8}$ we can
solve the resulting equation for $\psi^{(0)}$ at order $O(\alpha^{1/4})$: 
$\psi_{\sigma \sigma}^{(0)}+(1/\sigma)\psi^{(0)}-4 \sigma^2/\xi_0^6=0$.
After some algebra
we obtain the new approximation in normalized units:
$\psi=1+(P/4 \pi)-(P/8 \pi)r^2+(P/8\pi)^3 r^4/4$.
Then, the corresponding power-dependent refractive index perturbation reads:
\begin{equation}
  \label{eq:higherorder}
  \Delta n(R,\mathcal{P})=
  n_c [\frac{2\mathcal{P}}{\mathcal{P}_c}-
\frac{R^2}{R^2_c}(\frac{\mathcal{P}}{\mathcal{P}_c})^2+
\frac{R^4}{4 R_c^4}(\frac{\mathcal{P}}{\mathcal{P}_c})^3]\text{.}
\end{equation}
Thus higher-order approximations imply anarmonicity of the nonlocal potential, with a 
refractive index still depending on power. As the latter increases, higher powers of the ratio
 $\mathcal{P}/\mathcal{P}_c$ 
 must be taken into account, similar to local media with regards to the powers of the intensity.

Finally, comparing our theory with an actual experimental geometry, such as employed in \cite{Peccianti00}, typical parameters for a
$514nm$ wavelength and the E7 NLC (in SI units) are:
$n_a=1$, $K=10^{-11}$, $E=1.3\times10^{-4}$,$L=75\times10^{-6}$,
 $\Delta\epsilon_{RF}=20\,\epsilon_0$.
Correspondingly, $\mathcal{P}_c=2\times10^{-6}W$, 
$R_c=22\mu m$, and $\epsilon=10^{-6}$.
 Since $\mathcal{P}$ inside the cell was of the order of $0.1 mW$, 
we may state that in experiments the highly nonlocal regime ($\mathcal{P}>>\mathcal{P}_c$)
is being addressed. 
The soliton waist, after (\ref{eq:existencecurve}),
is of the order of $3\mu m$ and in agreement with the reported results.
Note that  $\Delta n(0,\mathcal{P})\cong5\times10^{-4}$, and the angle perturbation $\Psi$ of the order of $10^{-3}$ radians, thus justifying the
adopted model.

In conclusions, for the first time to our knowledge, we have presented a self-consistent 
analytical theory of two-dimensional spatial solitary waves in nonlocal media.
Our model has an intriguing unifying character, as it embraces several physical systems in which light self-trapping has recently been investigated. We believe that most of the observed SOS' in nematic liquid crystals are indeed \textit{accessible solitons}, in-as-much-as NLC are highly nonlocal. This shines new light on self-localization in liquid crystals.
Furthermore, we have presented the first derivation of a 
power-dependent constitutive relation for a real physical system, never
reported elsewhere. We confide that our results will stimulate new experiments towards a deeper
understanding of self-trapping in (highly) nonlocal nonlinear media and the development of novel all-optical devices. \cite{Peccianti02a}
%
\begin{acknowledgments}
C. C. aknowledges the Tronchetti Provera Foundation for a generous grant.
This work was supported by INFM (Advanced Research Project ``SPASONELIC").
\end{acknowledgments}

\end{document}